\documentclass[%
 reprint,
superscriptaddress,
 amsmath,amssymb,
 aps,
prb,
]{revtex4-2}

\usepackage{graphicx}
\usepackage{dcolumn}
\usepackage{bm}

\begin{document}

\preprint{APS/123-QED}

\title{Manipulating Topological Properties in Bi$_2$Se$_3$/BiSe/TMDC Heterostructures with Interface Charge Transfer}

\author{Xuance Jiang}

\affiliation{Department of Physics and Astronomy, Stony Brook University, Stony Brook, New York 11794, USA}
\affiliation{%
 Center for Functional Nanomaterials, Brookhaven National Laboratory, Upton, New York 11973, USA
}%

\author{Turgut Yilmaz}%
\affiliation{%
 National Synchrotron Light Source II, Brookhaven National Laboratory, Upton, New York 11973, USA
}%


\author{Elio Vescovo}
\affiliation{%
 National Synchrotron Light Source II, Brookhaven National Laboratory, Upton, New York 11973, USA
}%

\author{Deyu Lu}
\email{dlu@bnl.gov}
\affiliation{%
 Center for Functional Nanomaterials, Brookhaven National Laboratory, Upton, New York 11973, USA
}%


\date{\today}

\begin{abstract}
Heterostructures of topological insulator Bi$_2$Se$_3$ on transition metal dichalcogenides (TMDCs) offer a new materials platform for studying novel quantum states by exploiting the interplay among topological orders, charge orders and magnetic orders. The diverse interface attributes, such as material combination, charge re-arrangement, defect and strain, can be utilized to manipulate the quantum properties of this class of materials. Recent experiments of Bi$_2$Se$_3$/NbSe$_2$ heterostructures show signatures of strong Rashba band splitting due to the presence of a BiSe buffer layer, but the atomic level mechanism is not fully understood. We conduct first-principles studies of the Bi$_2$Se$_3$/BiSe/TMDC heterostructures  with five different TMDC substrates (1T phase VSe$_2$, MoSe$_2$, TiSe$_2$, and 2H phase NbSe$_2$, MoSe$_2$). We find significant charge transfer at both BiSe/TMDC and Bi$_2$Se$_3$/BiSe interfaces driven by the work function difference, which stabilizes the BiSe layer as an electron donor and creates interface dipole. The electric field of the interface dipole breaks the inversion symmetry in the Bi$_2$Se$_3$ layer, leading to the giant Rashba band splitting in two quintuple layers and the recovery of the Dirac point in three quintuple layers, with the latter otherwise only occurring in thicker samples with at least six  Bi$_2$Se$_3$ quintuple layers. Besides, we find that strain can significantly affect the charge transfer at the interfaces.  Our study presents a promising avenue for tuning topological properties in heterostructures of two-dimensional materials, with potential applications in quantum devices.

\end{abstract}

\maketitle


\section{\label{sec:intro} INTRODUCTION}

Topological insulators (TIs)~\cite{hasan2010colloquium,shen2012topological,moore2010birth,fu2007topological} are materials that are insulating in bulk but conducting on the surface. The Dirac surface states (DSSs) of TIs are protected by time-reversal symmetry and exhibit a characteristic spin-momentum locking with a linear dispersion relation.
These properties make TIs attractive for various applications in spintronics~\cite{fan2016spintronics}, quantum computing~\cite{he2019topological}, optoelectronics~\cite{plank2016opto,politano2017optoelectronic}, and thermoelectrics~\cite{xu2017topological,muchler2013topological}. The interplay among topological orders, charge orders (e.g. superconductivity and charge density wave) and magnetic orders can create exotic quantum states, which has drawn great interest in recent research~\cite{qi2011topological,fu2008superconducting,pan2022topological,mitsuishi2020switching,hor2010development,wang2016surface}.

\begin{figure}
\includegraphics[scale=0.2]{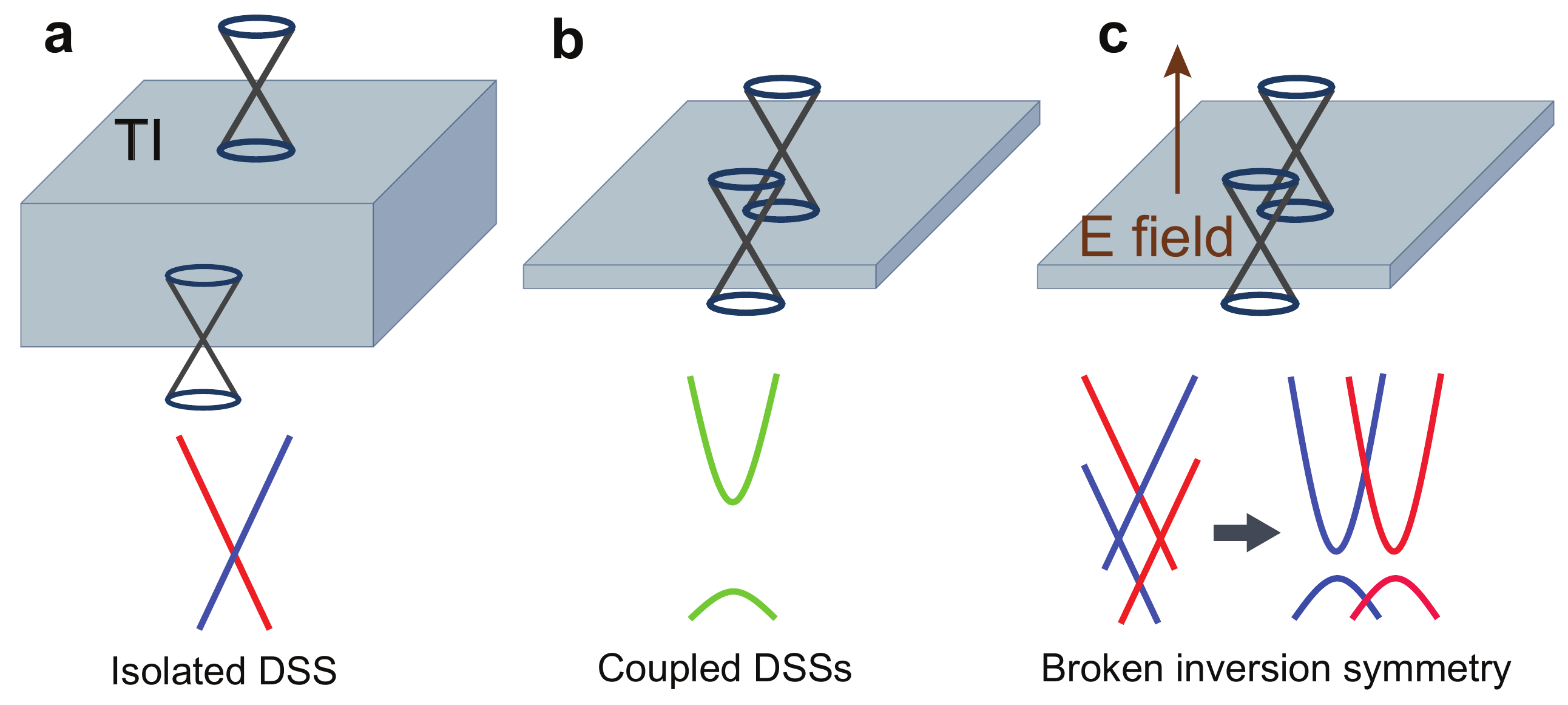}
\caption{\label{fig:dds} Schematic of topological insulators under the spatial confinement and external electric field (top) and the impact on the surface states (bottom). (a) Bulk topological insulator with two DSSs at the top and bottom surfaces. Red and blue lines indicate different spin polarizations. (b) Thin topological insulator, where two DSSs couple and open a gap. Green lines indicate the spin degenerate conduction and valence bands. (c) Thin topological insulator under an out-of-plane electric field. The bottom and top DSSs are separated in energy by the electric field without (left: two crosses) and with (right: gapped bands with large spin-orbit coupling) the DSSs coupling.}
\end{figure}

Heterostructures of two-dimensional (2D) materials or thin films bounded by van der Waals (vdW) forces are a fascinating playground for realizing new quantum materials. In particular, interface properties, such as the combination of materials, defects, strain and charge rearrangement, provide a wide, tunable design space to achieve new quantum properties~\cite{thiel2006tunable,he2016strain,lin20162d}. In this study, we demonstrate that interface charge transfer in vdW heterostructures can be used to tune DSSs of TIs, as shown in the schematics in Fig.~\ref{fig:dds}.

In bulk TIs, the top and bottom DSSs are degenerate in energy, but isolated in real space (Fig.~\ref{fig:dds}a). However, in thin film TIs, the DSSs can couple and open a gap~\cite{zhang2010crossover} (Fig.~\ref{fig:dds}b). In the bulk to 2D transition, the Dirac points and the corresponding spin-momentum locking are destroyed, which limits the application of thin film TIs in quantum information science. We propose to separate the top and bottom DSSs in thin film TIs in the energy domain by applying an out-of-plane electric field, which could arise from an interface dipole layer as a result of charge transfer. The electric field creates a potential energy offset in the two DSSs, which can still couple and open a gap. Two Dirac points re-emerge at different energies as the result of the inversion symmetry breaking and the strong Rashba spin-orbit coupling (SOC) in thin film TIs (Fig.~\ref{fig:dds}c).  
In this study, we focus on the realization of this mechanism using bismuth selenide (Bi$_2$Se$_3$) and transition metal dichalcogenide (TMDC) heterostructures.

Bi$_2$Se$_3$ is one of the most studied three-dimensional (3D) TIs~\cite{zhang2009topological,zhang2010first,mazumder2021brief}. It has a layered structure consisting of quintuple layers (QLs) of Se-Bi-Se-Bi-Se atoms that are weakly bounded by vdW forces (Figs.~\ref{fig:struct}a and~\ref{fig:struct}b). 
Zhang~\emph{et al.} reported that the DSS disappears when Bi$_2$Se$_3$ thin films have less than 6 QLs~\cite{zhang2010crossover}, as the top and bottom DSSs couple and open a gap.

TMDCs are two-dimensional materials that have diverse electronic and optical properties, such as semiconducting, metallic, or superconducting behaviors~\cite{manzeli20172d,choi2017recent,wang2017high,goli2012charge,yilmaz2023dirac}. Monolayer TMDCs have two types of structures: mirror symmetric 2H phase and inversion symmetric 1T phase as shown in Figs.~\ref{fig:struct}c-\ref{fig:struct}f. When Bi$_2$Se$_3$ is grown on TMDCs, novel topological properties can emerge depending on the interlayer coupling and the band alignment between the two materials.

Recent X-ray reflectivity (XRR) and X-ray diffraction (XRD) studies by Choffel~\emph{et al.} showed that when growing Bi$_2$Se$_3$ on metallic TMDCs, such as 2H phase of NbSe$_2$ (2H-NbSe$_2$) and 1T phase of TiSe$_2$ (1T-TiSe$_2$), VSe$_2$ (1T-VSe$_2$) and MoSe$_2$ (1T-MoSe$_2$), a buffer BiSe layer can form between them~\cite{choffel2021synthesis,choffel2021substituent}. Angle-resolved photoemission spectroscopy (ARPES) measurements revealed that the topological Dirac states reappear in the Bi$_2$Se$_3$/BiSe/NbSe$_2$ heterostructure, when the thickness of Bi$_2$Se$_3$ is only 3 QLs accompanied by a giant Rashba band splitting~\cite{yi2022crossover}, in contrast to the critical thickness of 6 QLs on SiC substrate~\cite{zhang2010crossover}. In addition, a giant Rashba splitting is found in Bi$_2$Se$_3$ bands~\cite{yi2022crossover}. 

It has been suggested that the formation of the BiSe buffer layer is responsible for the observed novel topological properties in the Bi$_2$Se$_3$/BiSe/NbSe$_2$ heterostructure. The BiSe buffer layer, similar to PbSe and LaSe, belongs to a rock-salt structured family that can form misfit layered compounds~\cite{kamminga2020misfit,mitchson2015antiphase,petvrivcek1993modulated,trump2014new,nader1997superconductivity,smeller2012structure,giang2010superconductivity,lin2010rational} with TMDCs, including 1T-TiSe$_2$, 2H-NbSe$_2$ and 1T-VSe$_2$. It has unique electronic properties such as one-dimensional electronic states~\cite{chikina2022one}. More importantly, the misfit structure suppresses interlayer bonding while allowing charge transfer to dope neighboring layers and tune quantum states in the heterostructure. For example, the charge transfer between misfit layers has a strong impact on the charge density wave and superconductivity phase~\cite{yao2018charge,song2022observation}. 

The structure of the BiSe monolayer is shown in Figs.~\ref{fig:struct}g and~\ref{fig:struct}h. The isolated system is metallic with occupied antibonding states and thus tends to donate electrons to adjacent materials as in misfit materials (BiSe)$_{1+\delta}$/NbSe$_2$~\cite{esters2018insights}. This property makes it a great candidate to modify the electronic properties of the interface. Besides, monolayer BiSe has a flexible structure that can grow on substrates with varying lattice constants. Scanning tunneling microscopy studies of the Bi$_2$Se$_3$/BiSe/NbSe$_2$ heterostructure show that the BiSe buffer layer has a distorted square lattice with the in-plane Se-Bi-Se bond angle ($\theta$ in Fig.~\ref{fig:struct}h) becoming either acute or obtuse~\cite{wang2014interface}. Density functional theory (DFT) calculations suggest that the isolated BiSe monolayer breaks the 4-fold rotation symmetry~\cite{mitchson2016structural} characterized by two different lattice parameters ($a$ and $b$) as shown in Fig.~\ref{fig:struct}g.

In spite of its intriguing topological properties, the atomic level structural details and the electronic structure of Bi$_2$Se$_3$/BiSe/TMDC heterostructures have not been fully understood. Due to the lattice mismatch in the three materials, a large supercell with hundreds of atoms is needed to minimize the artificial strain when building the heterostructure, which requires significant computational power. Existing \emph{ab initio} studies of Bi$_2$Se$_3$/BiSe/NbSe$_2$ heterostructures consider an incomplete structure model, which contains only the Bi$_2$Se$_3$/BiSe interface~\cite{yi2022crossover}. In order to qualitatively reproduce the Rashba splitting in Bi$_2$Se$_3$, this model requires a prematurely terminated relaxation to avoid unphysical structure distortion in the absence of the NbSe$_2$ layer~\cite{yi2022crossover}, since the NbSe$_2$ layer can play a crucial role in stabilizing the BiSe buffer layer. As such, a complete atomic scale physical picture of Bi$_2$Se$_3$/BiSe/TMDC heterostructures is still missing, and first-principles studies of the complete heterostructure are essential to gain insight into the atomic structure of the top and bottom interfaces, charge transfer characteristics, band alignment and topological properties of the Bi$_2$Se$_3$/BiSe/TMDC heterostructures.


In this study, we build a series of Bi$_2$Se$_3$/BiSe/TMDC models and conduct first-principles studies to reveal the origin of the giant Rashba splitting and Dirac crossing in few-QL Bi$_2$Se$_3$. We systematically investigate five hexagonal TMDC substrates (two 2H phase TMDCs: 2H-NbSe$_2$ and 2H-MoSe$_2$; three 1T phase TMDCs: 1T-TiSe$_2$, 1T-VSe$_2$ and 1T-MoSe$_2$; see Figs.~\ref{fig:struct}c-\ref{fig:struct}f) that have been reported in the experimental studies of new misfit compounds~\cite{choffel2021synthesis,choffel2021substituent} and Rahsba superconductivity pairing~\cite{yi2022crossover}. Due to the rich electronic properties of TMDCs and the tunable strong Rashba SOC induced in Bi$_2$Se$_3$, the Bi$_2$Se$_3$/BiSe/TMDC heterostructures are promising material platforms for future quantum applications, such as realizing topological superconductors and performing topological quantum computing.


The rest of the paper is organized as follows. In Section~\ref{sec:method}, we summarize the computational details. In Section~\ref{sec:bndalign}, we discuss the band alignment based on the work functions or electron affinity of heterostructure constituents. Then in Section~\ref{sec:crystal}, we analyze the structure and stability of the BiSe buffer layer. In Section~\ref{sec:chgtran}, we quantify the amount of charge transfer at the interfaces. In Section~\ref{sec:band}, we compare the calculated band structure of $n$QL-Bi$_2$Se$_3$/BiSe/NbSe$_2$ ($n$=1,2 and 3) with ARPES measurements. Finally, we investigate the strain effect in Section~\ref{sec:strain}.

\section{\label{sec:method} METHODS}

\begin{figure}
\includegraphics[width=3 in]{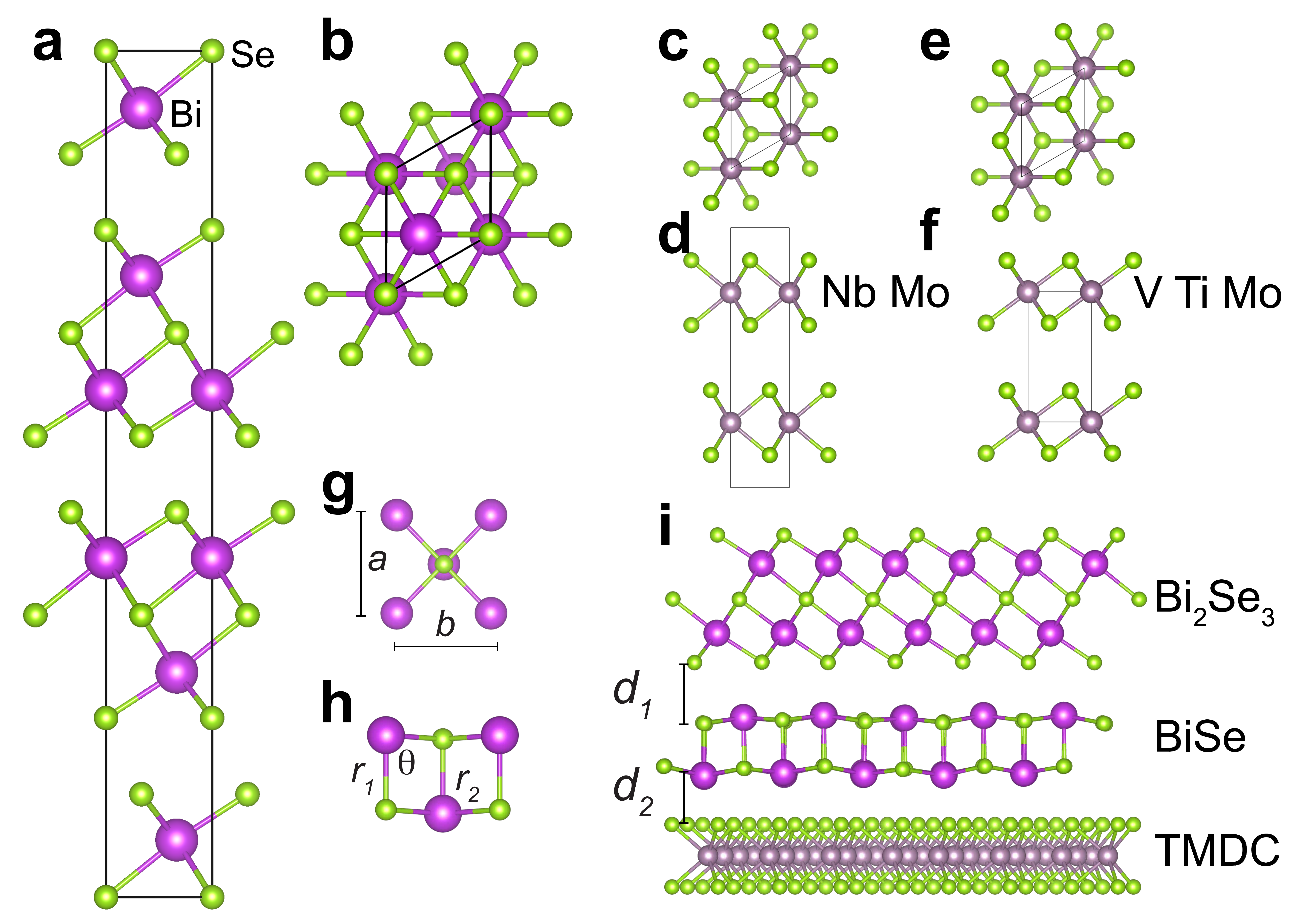} 
\caption{\label{fig:struct} (a) Side- and (b) top-view of the bulk Bi$_2$Se${_3}$ in the conventional cell indicated by the black rectangle. (c) Top- and (d) side-view of 2H phase TMDCs. (e) Top- and (f) side-view of 1T phase TMDCs. (g) Top- and (h) side-view of the BiSe monolayer. (i) Side view of the Bi$_2$Se$_3$/BiSe/TMDC heterostructure. The TMDC structure in the figure is taken from NbSe$_2$. $d_1$ and $d_2$ are inter-layer distances of the top and bottom interfaces. Purple: Bi; light green: Se;  red and green: transition metals.}
\end{figure}

\begin{table}
\caption{\label{tab:method} Supercells used in the Bi$_2$Se$_3$/BiSe/TMDC heterostructures.}
\begin{ruledtabular}
\begin{tabular}{ccccc}
 &\multicolumn{1}{c}{TMDC}&\multicolumn{1}{c}{BiSe}&\multicolumn{1}{c}{Bi$_2$Se$_3$}\\ \hline
 \\[-1em]
2H-NbSe$_2$&$\sqrt{13}\times\sqrt{39}\times1$& $3\times5\times1$  &$3\times3\sqrt3\times1$ \\ \hline
 \\[-1em]
1T-TiSe$_2$&$7\times7\times1$&  $\sqrt{34}\times\sqrt{34}\times1$  & $6\times6\times1$ \\ \hline
 \\[-1em]
XSe$_2$\footnote{XSe$_2$ represents 1T-VSe$_2$ and 1T or 2H phase MoSe$_2$.} & $\sqrt{57}\times\sqrt{57}\times1$&  $\sqrt{34}\times\sqrt{34}\times1$  & $6\times6\times1$ \\ 
\end{tabular}
\end{ruledtabular}
\end{table}

DFT calculations are performed using the projector-augmented-wave (PAW) method~\cite{kresse1999ultrasoft} implemented in the Vienna Ab initio Simulation Package (VASP)~\cite{kresse1993ab,kresse1996efficiency,kresse1996efficient}. The exchange-correlation effects are treated under the generalized gradient approximation using the Perdew-Burke-Ernzerhof (PBE) functional~\cite{perdew1996generalized}. The non-local vdW interaction is modeled by the vdW-D2 method~\cite{grimme2006semiempirical}.

Due to the lattice mismatch at the TMDC/BiSe and BiSe/Bi$_2$Se$_3$ interfaces, slab models of the heterostructures are built from supercells with 300 to 600 atoms as summarized in Table~\ref{tab:method} (see also Fig.~\ref{fig:struct}i and Supplemental Material~\cite{supp}). The heterostructure models contain a TMDC monolayer, a BiSe monolayer and one to three QLs of Bi$_2$Se$_3$. Except for explicitly mentioned, 2QL-Bi$_2$Se$_3$ is used throughout this study. The lattice mismatch in the supercell is less than 1.5\% between TMDC and Bi$_2$Se$_3$ and less than 4\% between BiSe and Bi$_2$Se$_3$. 
To avoid the spurious interaction between periodic images of the slab, we include a vacuum region of at least 20 Å and apply the dipole correction~\cite{bengtsson1999dipole} in the DFT calculations. The structure relaxation is performed by fixing the monolayer TMDC substrate and relaxing the rest of the system until the total energy difference is less than $10^{-5}$ eV and the force is less than $0.01$ eV/\AA.


To study the interface charge transfer, the electron density difference is calculated by subtracting the electron densities of isolated TMDCs, BiSe and Bi$_2$Se$_3$ from that of the heterostructure. The iso-surface plots are generated with the VASPKIT package~\cite{VASPKIT}. 
The band structures of the heterostructures are computed including SOC and unfolded to the Bi$_2$Se$_3$ Brillouin zone using the Vaspbandunfolding package~\cite{vaspbandunfolding}. 

\section{\label{sec:result}Results and Discussion}

\subsection{\label{sec:bndalign} Band alignment of the Bi$_2$Se$_3$/BiSe/TMDC heterostructures}
The free-standing BiSe monolayer (see Figs.~\ref{fig:struct}g and~\ref{fig:struct}h) is unstable, because of the strong Coulomb repulsion from the excess electrons on the Bi atom (one per Bi atom). Our Bader charge analysis~\cite{tang2009grid} shows that each Bi atom in BiSe has 0.3 electrons more than that in bulk Bi$_2$Se$_3$. As a result, the isolated BiSe monolayer favors the zigzag shape in the lateral direction rather than the rectangular shape (as shown from the side view in Fig.~\ref{fig:struct}h), in order to maximize the Bi-Bi distance, with the Se-Bi-Se bond angle $\theta=87^{\circ} $ and the out-of-plane BiSe bond length $r_1= r_2=2.92$ \AA. Due to the breaking of the 4-fold rotation symmetry~\cite{mitchson2015antiphase}, in-plane parameters ($a=4.44$~{\AA} and $b=4.13$~{\AA}) have different values (see Fig.~\ref{fig:struct}g). 
One practical strategy to stabilize the BiSe monolayer is to put it in contact with electron acceptors forming heterostructures~\cite{esters2018insights}. Since the energy level alignment at the interface plays a critical role in determining the charge re-arrangement and the thermodynamic stability of the heterostructures, below we investigate the energy level alignment at the BiSe/TMDC and Bi$_2$Se$_3$/BiSe interfaces.

We calculate the work function ($W$) of the isolated monolayer BiSe and metallic monolayer TMDCs, as well as the electron affinity ($E_a$) of the isolated semiconducting monolayer TMDC and bulk Bi$_2$Se$_3$ as shown in Table~\ref{tab:table1}. At the PBE level, $W_\mathrm{BiSe}=3.72$ eV, which is lower than that of all the metallic monolayer TMDCs, 1T-MoSe$_2$ (4.52 eV), 1T-VSe$_2$ (5.01 eV), 1T-TiSe$_2$ (5.31 eV) and 2H-NbSe$_2$ (5.52 eV), and $E_a$ of the semiconducting 2H-MoSe$_2$ monolayer (3.90 eV). Our results show good agreement with the PBE calculations in the literature, e.g. 2H-NbSe$_2$ (5.54 eV~\cite{ni2022organic}), 1T-TiSe$_2$ (5.35 eV~\cite{liu2016van}), 1T-VSe$_2$ (5.05 eV~\cite{liu2016van}), 1T-MoSe$_2$ (4.58 eV~\cite{liu2016van}), 2H-MoSe$_2$ (3.86 eV~\cite{kim2021thickness}). They are in reasonable agreement with measured $W$ of the 1T-VSe$_2$ monolayer (5.0~\cite{wong2019evidence}, 5.52 eV
~\cite{liu2018epitaxially}) and $E_a$ of the semiconducting 2H-MoSe$_2$ monolayer (3.5, 3.8 eV~\cite{zhang2019band}).

Based on Table~\ref{tab:table1}, at the BiSe/TMDC interface, $W_{\text{TMDC}}$ or $E_{a}^{\text{TMDC}}$ is larger than $W_{\text{BiSe}}$ by up to 1.80 eV, which drives electron transfer from BiSe to TMDCs. At the Bi$_2$Se$_3$/BiSe interface,  $W_{\text{BiSe}}$ is lower than $E_{a}^{\text{Bi}_2\text{Se}_3}$ by 1.68 eV. Therefore, BiSe also donates electrons to Be$_2$Se$_3$. The charge transfer at both the top and bottom interfaces can significantly reduce the Coulomb repulsion of excess electrons in BiSe, which in turn can stabilize the BiSe buffer layer. 

\begin{figure}
\includegraphics[scale=0.24]{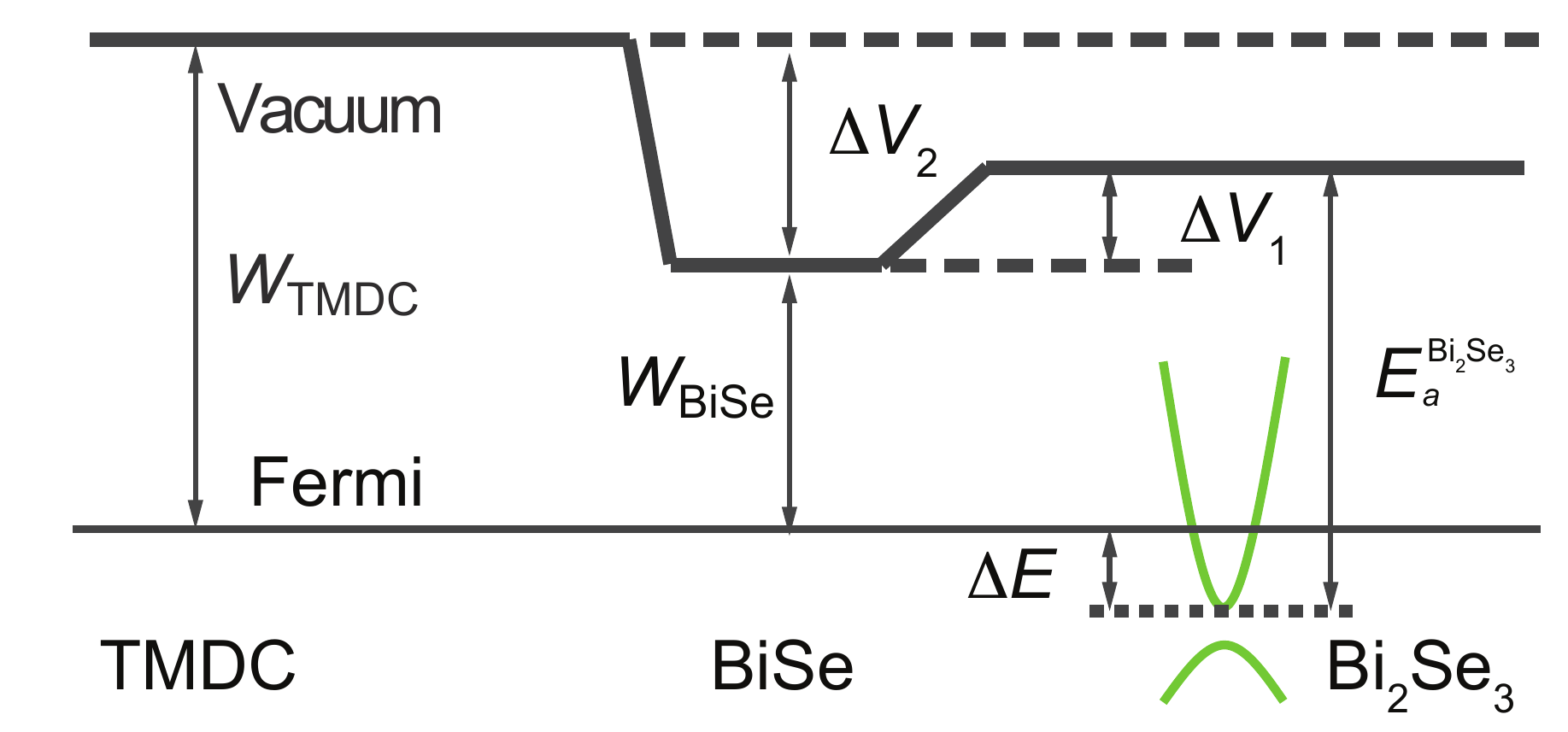}
\caption{\label{fig:bndalign} Band alignment diagram of Bi$_2$Se$_3$/BiSe/TMDC heterostructures. The thick solid line, thin solid line and green curves represent the vacuum level, Fermi level, and Bi$_2$Se$_3$ bands, respectively. } 
\end{figure}

The charge transfer in the Bi$_2$Se$_3$/BiSe/TMDC heterostructures creates interface dipoles at both the top and bottom interfaces. The resulting dipole field shifts the vacuum potential, which determines the overall band alignment in the heterostructures. A qualitative picture is outlined in Fig.~\ref{fig:bndalign}.  As illustrated in this diagram, the band alignment at the  BiSe/TMDC and Bi$_2$Se$_3$/BiSe interfaces is dominated by the dipole potential $\Delta V_2$ and $\Delta V_1$, respectively. The relative position of the conduction band minimum (CBM) of Bi$_2$Se$_3$ ($\Delta E$) against the Fermi level is associated with $\Delta V_1$  through $\Delta E=W_{\mathrm{BiSe}}-E_{a}^{\text{Bi}_2\text{Se}_3} -\Delta V_1$.
We plot the Hartree potential of the 2QL-Bi$_2$Se$_3$/BiSe/NbSe$_2$ heterostructure in Fig.~S3 of the Supplemental Material~\cite{supp}. The vacuum level difference of 0.33 eV between the bottom and top of the heterostructure corresponds to $\Delta V_2-\Delta V_1$, as the net effect of the interface dipole. From the projected density of states (PDOS) of the  Bi$_2$Se$_3$ bands, $\Delta E$ is estimated to be 0.44 eV.
We note that $W$ and $E_a$ in the heterostructure (as shown in Fig.~\ref{fig:bndalign}) can be slightly different from those of the isolated system due to structure distortion and charge re-arrangement. For example, in the 2QL-Bi$_2$Se$_3$/BiSe/NbSe$_2$, Bi$_2$Se$_3$ has a slightly larger $E_{a}^{\text{Bi}_2\text{Se}_3}$ than the free-standing system by 3\%. 



\begin{table*}
\caption{\label{tab:table1}Work functions (eV) of the metallic monolayer TMDCs (2H-NbSe$_2$, 1T-TiSe$_2$, 1T-VSe$_2$, 1T-MoSe$_2$) and BiSe; electron affinities (eV) of the semiconducting monolayer TMDC (2H-MoSe$_2$) and 2QL-Bi$_2$Se$_3$. Values in the parenthesis are taken from the literature.}
\begin{ruledtabular}
\begin{tabular}{cccccccc}
 &\multicolumn{5}{c}{TMDCs}&\multicolumn{1}{c}{BiSe}&\multicolumn{1}{c}{Bi$_2$Se$_3$}\\
  &2H-NbSe$_2$&1T-TiSe$_2$&1T-VSe$_2$&1T-MoSe$_2$&2H-MoSe$_2$&&\\ \hline
 DFT &5.52 (5.54\footnote{\label{dft1}See Ref.~\cite{ni2022organic}})&5.31(5.35\footnote{\label{bulk}See Ref.~\cite{liu2016van}}) &5.01(5.05\footref{bulk})&4.52(4.58\footref{bulk})&3.90 (3.86\footnote{\label{4foot}See Ref.~\cite{kim2021thickness}})&3.72&5.40 \\ \hline
  experiment&-&-&5.0\footnote{See Ref.~\cite{wong2019evidence}},5.52\footnote{\label{vse2}See Ref.~\cite{liu2018epitaxially}}&-&3.5,3.8\footnote{\label{5foot}(3.8 ± 0.1) eV on Al$_2$O$_3$/Si and (3.5 ± 0.1) eV on SiO$_2$/Si. See Ref.~\cite{zhang2019band}}&-&- \\ 
 
\end{tabular}
\end{ruledtabular}
\end{table*}

\begin{table*}
\caption{\label{tab:struct} Structural parameters of the relaxed 2QL-Bi$_2$Se$_3$/BiSe/TMDC heterostructures ($r_1$, $r_2$, $d_1$ and $d_2$ in \AA) and the adhesive energy of the BiSe/TMDC interface ($\varepsilon_{\mathrm{BiSe}/\mathrm{TMDC}}^{adh}$ in eV). $r_1$ and $r_2$ are the average out-of-plane Bi-Se bond length in the BiSe monolayer. In the free-standing BiSe monolayer, $r_1=r_2=2.92$~{\AA}. $d_1$ ($d_2$) is the interlayer distance at the top (bottom) interface.}
\begin{ruledtabular}
\begin{tabular}{cccccc}
TMDC&2H-NbSe$_2$&1T-TiSe$_2$&1T-VSe$_2$&1T-MoSe$_2$&2H-MoSe$_2$\\ \hline

$r_1$ &2.75&2.79&2.77&2.80&2.82 \\\hline
$r_2$ &2.84&2.84&2.82&2.82&2.75 \\\hline

$d_1$ &3.08&3.06&3.08&3.05&2.98 \\\hline
$d_2$ &2.92&3.06&3.12&3.15&3.27 \\\hline

$\varepsilon_{\mathrm{BiSe}/\mathrm{TMDC}}^{adh}$&-3.98&-2.80&-3.03&-2.82&-1.82 \\

\end{tabular}
\end{ruledtabular}
\end{table*}

\subsection{\label{sec:crystal}Structure and stability of the BiSe buffer layer}

Compared to the free-standing BiSe monolayer, the out-of-plane Bi-Se bond lengths ($r_1$ and $r_2$) are about 5\% shorter in the fully relaxed heterostructures (See Fig.~\ref{fig:struct}i and Table \ref{tab:struct}). Unlike other misfit structures~\cite{esters2018insights},  the BiSe monolayer in Bi$_2$Se$_3$/BiSe/TMDC heterostructures can have significant structure distortion (i.e., $r_1\ne r_2$), which breaks the inversion symmetry. The degree of this distortion depends on the relative energy level alignment on the top and bottom interfaces. At the top interface, $\Delta_T=E_{a}^{\text{Bi}_2\text{Se}_3}-W_{\text{BiSe}}=1.68$ eV is the driving force of electron donation from BiSe to Bi$_2$Se$_3$. Similarly, at the bottom interface, $\Delta_B=W_{\text{TMDC}}(E_{a}^{\text{TMDC}})-W_{\text{BiSe}}$ causes electron to flow from BiSe to TMDC. Larger charge transfer results in stronger coupling at the interface, which will pull Bi atoms out of the Se plane in BiSe. Therefore, the ratio $\eta=\Delta_B/\Delta_T$ can be used to qualitatively determine the degree of asymmetry of the BiSe monolayer. For example, in Bi$_2$Se$_3$/BiSe/NbSe$_2$, $\eta$ has the largest value of 1.07. As a result, Bi atoms in the top surface of BiSe move towards Bi$_2$Se$_3$ by 0.28 \hbox{\AA} relative to the top Se plane, while Bi atoms in the bottom surface move towards NbSe$_2$ by 0.38 \hbox{\AA} relative to the bottom Se plane. Consequently, $r_2=2.84$ \hbox{\AA} is longer than $r_1= 2.75$ \hbox{\AA} by 3\%. In Bi$_2$Se$_3$/BiSe/1T-MoSe$_2$ with a moderate $\eta$ of 0.48, the BiSe layer is almost inversion symmetric, as $r_2$ is longer than $r_1$ by only 0.7\%. The trend is completely reversed in Bi$_2$Se$_3$/BiSe/2H-MoSe$_2$ with a very small $\eta$ of 0.11, where $r_2=2.75$~{\AA} is shorter than $r_1=2.82$~{\AA} by 2.5\%.

The coupling strength at the interface can also be characterized by inter-layer distances $d_1$ and $d_2$ at the top and bottom interfaces, respectively, as shown in Fig.~\ref{fig:struct}i. From Table \ref{tab:struct}, we can see that as $\Delta_B$ decreases, $d_2$ increases from 2.92 to 3.27 \hbox{\AA} implying weaker interactions. On the other hand, $d_1$ shows a mild opposite trend, decreasing from 3.08 to 2.98~\AA. This suggests that stronger interaction at the BiSe/TMDC interface slightly weakens the interaction at the Bi$_2$Se$_3$/BiSe interface. This trend is consistent with detailed charge transfer analysis in Section~\ref{sec:chgtran}.

In the bottom-up growth of heterostructures (e.g. using molecular-beam epitaxy), the adhesive energy of the BiSe/TMDC interface is a crucial factor of the thermodynamic stability.
We calculate the adhesive energy for each TMDC substrate,
\begin{equation}
    \varepsilon^{\mathrm{BiSe}/\mathrm{TMDC}}_{adh}=\left(E_{\mathrm{BiSe}/\mathrm{TMDC}}-n\,\varepsilon_{\mathrm{BiSe}}-m\,\varepsilon_{\mathrm{TMDC}}\right),
\end{equation}
where $E_{\mathrm{BiSe}/\mathrm{TMDC}}$, $\varepsilon_{\mathrm{BiSe}}$ and $\varepsilon_{\mathrm{TMDC}}$ are the energies of the $\mathrm{BiSe}/\mathrm{TMDC}$ interface and individual components per unit cell, respectively. $n$ and $m$ are the corresponding numbers of BiSe and TMDC 2D unit cells in the heterostructure supercell. The results in Table~\ref{tab:struct} show that as $W_{\mathrm{TMDC}}$ increases, $\varepsilon^{\mathrm{BiSe}/\mathrm{TMDC}}_{adh}$ becomes more negative from -2.82 eV (1T-MoSe$_2$) to -3.98 eV (2H-NbSe$_2$) per BiSe chemical formula, as a result of stronger charge transfer and coupling between BiSe and TMDC substrates. However, for the semiconducting 2H-MoSe$_2$ substrate, $\Delta_B=0.18$ eV is very small. This implies a much weaker charge transfer, and the resulting $\varepsilon^{\mathrm{BiSe}/\mathrm{TMDC}}_{adh}=-1.82$ eV is the smallest among five TMDCs. Our adhesive energy calculations suggest that a BiSe buffer layer can preferably grow on metal TMDC substrates with large work functions, which is in line with experimental observations~\cite{choffel2021synthesis}.

\begin{figure}
\includegraphics[scale=0.25]{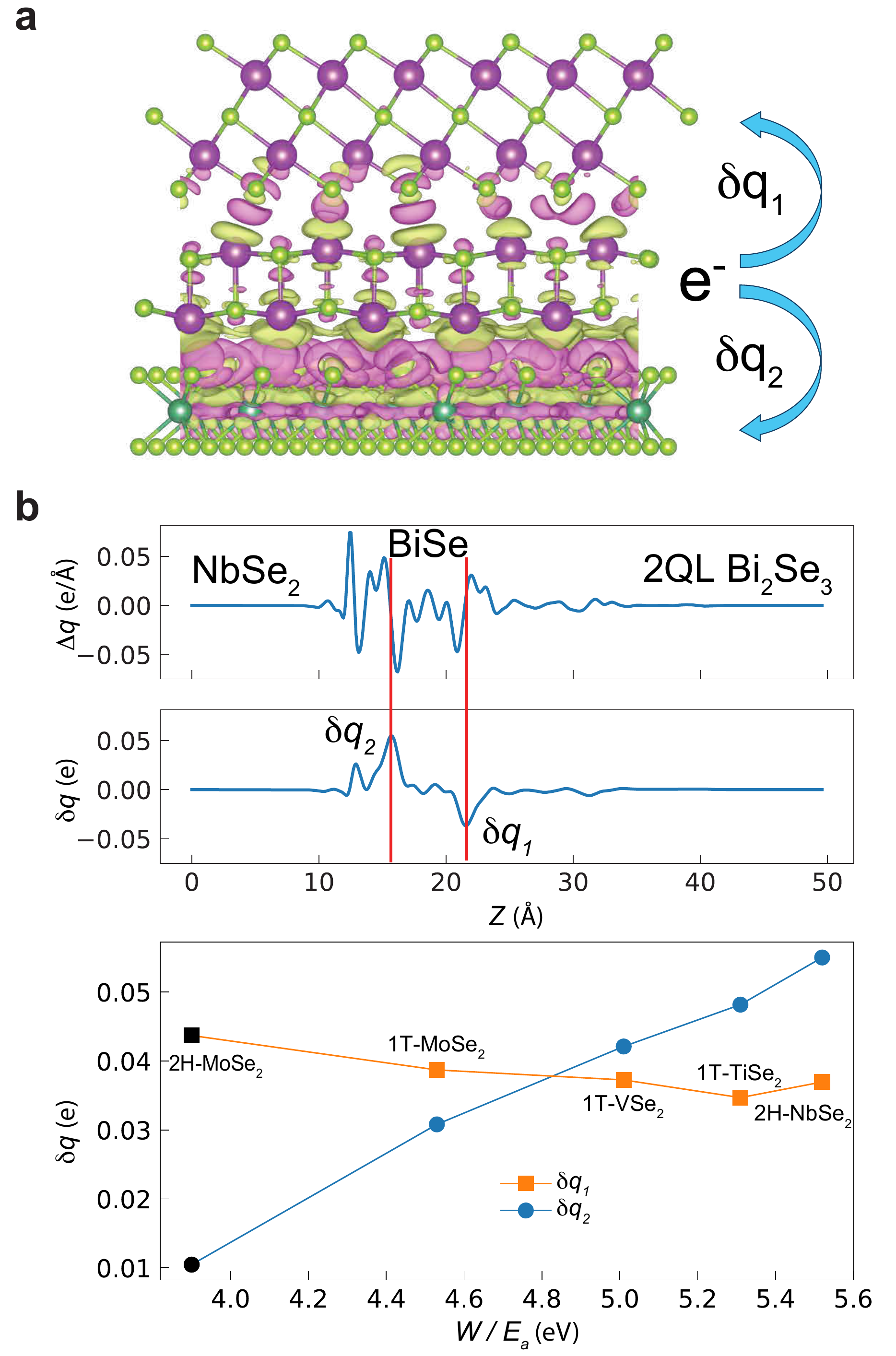}
\caption{\label{fig:chg} (a) Isosurface plot of electron density difference of heterostructure 2QL-Bi$_2$Se$_3$/BiSe/NbSe$_2$. For clarity, only the bottom QL of Bi$_2$Se$_3$ is shown.  Pink and yellow regions indicate the electron gain and loss, respectively. (b) Top: the plane averaged electron density curve of 2QL-Bi$_2$Se$_3$/BiSe/NbSe$_2$; Bottom: the corresponding cumulative charge profile in the unit of the electron per BiSe formula. Two peaks indicated by the red lines correspond to the amount of charge transfer at two interfaces. (c) Effect of the work function or electron affinity of TMDC substrates on interface charge transfer. The black symbols correspond to the semiconducting 2H-MoSe$_2$.} 
\end{figure}

\subsection{\label{sec:chgtran} Interface charge transfer}
We examine the electronic properties of two interfaces in Bi$_2$Se$_3$/BiSe/TMDC and determine the amount of charge transfer. In Fig.~\ref{fig:chg}a, we plot the isosurface of the electron density difference between the combined system and its individual components. One can clearly see that electron flows out of Bi $p_z$ orbitals in BiSe (yellow blobs) into Se $p_z$ orbitals in either TMDC or Bi$_2$Se$_3$ (pink blobs).


The plane-averaged electron density difference plot of the heterostructure is shown in Fig.~\ref{fig:chg}b (top), where significant charge transfer occurs from the BiSe buffer layer (negative regions) to both NbSe$_2$ and Bi$_2$Se$_3$ (positive regions). To quantify the amount of charge transfer at the two interfaces, we integrate the plane-averaged electron density difference along the surface normal direction to obtain the cumulative charge profile, as shown in Fig.~\ref{fig:chg}b (bottom). Two pronounced peaks are indicated by the vertical red lines, corresponding to the net amount of charge transfer on each interface. The positive peak at the bottom interface corresponds to electron donation to NbSe$_2$ with $\delta q_2=0.06$ $e^-$ and the negative peak at the top interface corresponds to electron donation to Bi$_2$Se$_3$ with $\delta q_1=0.04$ $e^-$. 

The trend of the charge transfer with respect to $W$ ($E_a$) is shown in Fig.~\ref{fig:chg}c. As we expect, $\delta q_2$ grows monotonously with $W$ ($E_a$), varying from 0.01 to 0.06 $e^-$. The smallest $\delta q_2$ value of 0.01 $e^-$ is obtained on the 2H-MoSe$_2$ substrate, with $W=0.18$ eV. 
On the other hand, $\delta q_1$ overall shows a mild decay trend against $W$ ($E_a$), except for 2H-NbSe$_2$ which has a slightly larger $\delta q_1$ (0.037 $e^-$) than 1T-TiSe$_2$ (0.035 $e^-$). Since $\delta q_1$ is primarily determined by the constant value of $\Delta_T$ (1.68 eV), it exhibits a much narrower dynamic range from 0.035 to 0.044. Another factor that may contribute to the non-monotonic behavior of $\delta q_1$ is the in-plane strain in the heterostructure, which will be discussed in Section \ref{sec:strain}. There is more tensile strain in the Bi$_2$Se$_3$ and BiSe layers in Bi$_2$Se$_3$/BiSe/NbSe$_2$ heterostructure than that in Bi$_2$Se$_3$/BiSe/TiSe$_2$ (see Supplemental Material~\cite{supp}), which may cause an enhancement of $\delta q_1$ in the former and a reduction of $\delta q_1$ in the latter. 

\begin{figure*}
\includegraphics[scale=1.8]{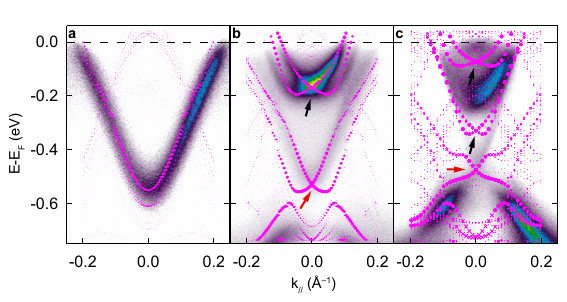}
\caption{\label{fig:bnd2ql} Comparison between experimental~\cite{yi2022crossover} and DFT band structures of (a) 1QL-, (b) 2QL- and (c) 3QL-Bi$_2$Se$_3$/BiSe/NbSe$_2$ heterostructures. The pink circles are unfolded band structures from DFT.  Black arrows indicate the QW states; the red arrow indicates the Dirac point.  Bottom and top QL states in (c) are indicated by  crosses and circles, respectively.}
\end{figure*}

\subsection{\label{sec:band}Band structure and giant Rashba splitting}
Recent ARPES measurements  revealed interesting topological properties in few-layer Bi$_2$Se$_3$/BiSe/NbSe$_2$ heterostructures~\cite{yi2022crossover}. To understand the ARPES data, the calculated band structures (pink dots) of the 1QL-, 2QL- and 3QL-Bi$_2$Se$_3$/BiSe/1ML-NbSe$_2$ heterostructures are unfolded to the Brillouin zone of Bi$_2$Se$_3$ and overlaid on top of the measured band structure from Ref.~\cite{yi2022crossover} in Fig.~\ref{fig:bnd2ql}. For this comparison, the calculated Fermi levels are slightly shifted by up to 0.15 eV to match the experiment. The amount of this adjustment is comparable to the differences in work function between DFT and experiment as shown in Table~\ref{tab:table1}.

 


In the 1QL-Bi$_2$Se$_3$/BiSe/NbSe$_2$ band structure (see Fig.~\ref{fig:bnd2ql}a), ARPES shows a V-shape conduction band derived from Bi and Se $p$ orbitals of Bi$_2$Se$_3$, which is reproduced very well by our DFT calculations. The Bi$_2$Se$_3$ CBM is lower than the Fermi level by 0.55 eV, which is qualitatively reproduced by 0.70 eV from DFT. The valence band of Bi$_2$Se$_3$ is at lower energies outside the range of Fig.~\ref{fig:bnd2ql}a.


In 2QL-Bi$_2$Se$_3$/BiSe/NbSe$_2$, the surface states in Bi$_2$Se$_3$ are gapped, and the CBM shows a band splitting near the $\Gamma$ point indicated by the red arrow (see Fig.~\ref{fig:bnd2ql}b). Due to the charge transfer between BiSe and Bi$_2$Se$_3$, the negative interface dipole along the out-of-plane direction breaks the inversion symmetry, which separates the DSSs in energy and weakens their coupling. DFT band structure qualitatively reproduces both the band gap and band splitting at the CBM, consistent with the schematic diagram in Fig.~\ref{fig:dds}c. The underestimated gap and overestimated surface band splitting in DFT, as compared to experiment, are likely due to the limitation of the DFT method, which may not capture the screened long-range Coulomb repulsion between top and bottom surface states accurately~\cite{liu2023surface}. 


Notably, we observe the giant Rashba band splitting in the quantum well (QW) state indicated by the black arrow, which is a strong evidence of the dipole-induced giant Rashba SOC effect. In contrast, the QW state in the Bi$_2$Se$_3$/NbSe$_2$ band structure (i.e., without the BiSe buffer layer) resembles the pristine 2QL-Bi$_2$Se$_3$ with negligible Rashba splitting  (see Fig. S2 in the Supplemental Material~\cite{supp}). The Rashba constant is defined as $\alpha_R = 2\Delta E/\Delta k$, where $\Delta E$ and $\Delta k$ are energy and momentum splitting values. Based on the ARPES measurement, $\Delta E$ = 0.019 eV and $\Delta k$ = 0.038 \AA$^{-1}$, which yield $\alpha_R$ = 1.0 eV$\cdot$\AA~\cite{yi2022crossover}. Values of $\Delta E$, $\Delta k$ and $\alpha_R$ from simulated band structures of 2QL-Bi$_2$Se$_3$/BiSe/NbSe$_2$ heterostructures are listed in Table \ref{tab:table3}. In the heterostructure without strain, $\alpha_R= 1.1$ eV$\cdot$\AA~is in excellent agreement with experiment. Note that the Au(111) surface, as a typical giant Rashba SOC material, has a Rashba constant of 0.33 eV$\cdot$\AA~\cite{lashell1996spin}, and bulk BiTeI, among the highest Rashba SOC materials, has a Rashba constant of 3.8 eV$\cdot$\AA~\cite{ishizaka2011giant}. 

When the thickness of Bi$_2$Se$_3$ in the heterostructure increases to 3QLs, the Dirac point re-appears as indicated by the red arrow (see Fig.~\ref{fig:bnd2ql}c), which is clearly reproduced by DFT. This feature is due to both the spatial separation of the top and bottom surface states and their energy offset caused by the interface dipole field. In fact, the interface electric field pulls the bottom surface state down into the bulk states, so the coupling between top and bottom surface states is weaker than the free-standing 3QL-Bi$_2$Se$_3$. In contrast, in pristine Bi$_2$Se$_3$ thin films the surface Dirac point appears only when the thickness is equal to or more than 6 QLs. In addition to the surface Dirac point, at higher energies, ARPES data also exhibit two QW states with giant Rashba band splitting. These QW states are reproduced by DFT calculations (indicated by black arrows), while the energy positions in DFT are closer to the Dirac point than the experiment by 0.09 eV, likely due to the limitation of the PBE functional. In Fig.~\ref{fig:bnd2ql}c, we distinguish the quasiparticle states dominated by the bottom and top QLs with crosses and dots, respectively. The bottom-QL-dominated states do not appear in the experimental band structure, likely due to the thickness sensitivity of the ARPES measurement.

Overall, our calculations reproduce well all the key topological features in the ARPES measurement of few-layer Bi$_2$Se$_3$/BiSe/NbSe$_2$ heterostructures, and the calculated giant Rashba constant is in excellent agreement with experiment.

In Table~\ref{tab:table3}, we compare $\alpha_R$ of different TMDC heterostructures based on the band splitting of the first QW state in 2QL-Bi$_2$Se$_3$/BiSe/TMDC. The calculated $\alpha_R$ are between 1.0 eV$\cdot${\AA} (1T-TiSe$_2$) and 1.2 eV$\cdot${\AA} (2H-MoSe$_2$). Note that $\alpha_R$ of the 2H-MoSe$_2$ heterostructure is 9\% larger than that of the 2H-NbSe$_2$ heterostructure, which is consistent with the charge transfer analysis, as larger $\delta q_1$ induces larger Rashba splitting.

\begin{table*}
\caption{\label{tab:table3}Rashba strength parameter of different 2QL-Bi$_2$Se$_3$/BiSe/TMDC heterostructures.}
\begin{ruledtabular}
\begin{tabular}{ccccccc}
 &\multicolumn{5}{c}{TMDCs}&\multicolumn{1}{c}{Exp\footnote{See Ref.~\cite{yi2022crossover}}}\\
 &2H-NbSe$_2$&1H-TiSe$_2$&1T-VSe$_2$&1T-MoSe$_2$&2H-MoSe$_2$&2H-NbSe$_2$ \\ \hline
$\alpha_R$ (eV$\cdot$\AA)  &1.1&1.0&1.1&1.1&1.2&1.0\\ 
\end{tabular}
\end{ruledtabular}
\end{table*}

\begin{figure}
    \centering
    \includegraphics[width=1\linewidth]{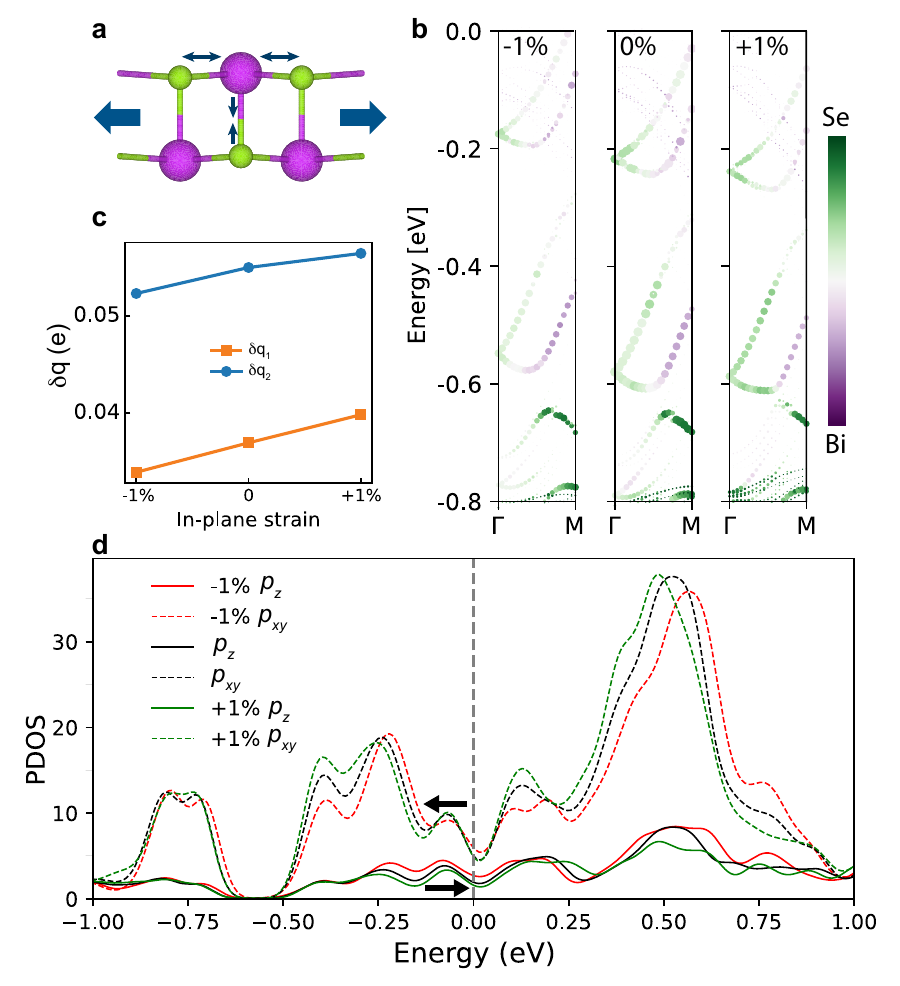}
    \caption{(a) Illustration of the structural change of the BiSe layer in the heterostructure under the tensile strain. The large blue arrows indicate the in-plane tensile strain. The small arrows indicate the change in the bond length. (b) Band structure of Bi$_2$Se$_3$/BiSe/NbSe$_2$ under 1\% compressive (left), zero (middle) and +1\% tensile (right) strain projected onto Bi$_2$Se$_3$ orbitals. Pink and green colors are Bi and Se dominated orbitals. (c) The effect of in-plane strain on the charge transfer of the Bi$_2$Se$_3$/BiSe/NbSe$_2$ heterostructure. (d) Projected density of states of the BiSe layer under different amount of in-plane tensile strain. Black arrows highlight the change of the projected density of states.}
    \label{fig:pdos}
\end{figure}

\subsection{\label{sec:strain}Effect of strain}
The dipole field from the Bi$_2$Se$_3$/BiSe interface is the critical factor that leads to the novel topological properties in the Bi$_2$Se$_3$/BiSe/TMDC heterostructures. Therefore it is advantageous to have the ability to tune $\delta q_1$ in order to enhance this dipole field, combined with a large enough $\delta q_2$ to stabilize the BiSe buffer layer. As shown in Fig.~\ref{fig:chg}c, the range of $\delta q_1$ is about 0.0027 $e^-$ among the TMDCs (between 1T-MoSe$_2$ that has a small work function and 2H-NbSe$_2$ that has a large work function) that can grow the BiSe layer in experiment~\cite{choffel2021synthesis}. Here we consider strain as another means to control the interface charge transfer.

Strain can have significant effects on the electronic properties of 2D materials. Specifically, we investigate the effect of strain on the charge transfer at the two interfaces. We use the 2QL-Bi$_2$Se$_3$/BiSe/NbSe$_2$ heterostructure as an exemplary system and apply a uniform in-plane strain between -1\% and +1\% to mimic the strain induced by the lattice mismatch between TMDCs and their substrate. We find that tensile strain enhances the charge transfer at the Bi$_2$Se$_3$/BiSe interface, while compressive strain suppresses it as shown in Fig.~\ref{fig:pdos}c. In fact, 1\% tensile strain increases $\delta q_1$ by 0.0029 $e^-$, which is slightly larger than the effect (0.0027 $e^-$) caused by different TMDCs. As a result, heterostructures under tensile strain have larger Rashba splitting as shown in Fig.~\ref{fig:pdos}b. The calculated $\alpha_R$ increases as a function of strain: $\alpha_R=1.0$, 1.1 and 1.4 eV$\cdot${\AA} under 1\% compressive strain, zero strain and 1\% tensile strain, respectively. On the other hand, 1\% tensile strain increases $\delta q_2$ by 0.0015 $e^-$, which is much smaller than the effects of different TMDCs (0.024 $e^-$).

To understand the effects of strain on charge transfer, we focus on the BiSe buffer layer and plot the PDOS of BiSe in Fig.~\ref{fig:pdos}d. The in-plane tensile strain effectively stretches the in-plane Bi-Se bond and compresses the out-of-plane Bi-Se bond in the BiSe monolayer as shown in Fig.~\ref{fig:pdos}a. Consequently, the $p_x$/$p_y$ anti-bonding orbitals move to lower energy and $p_z$ anti-bonding orbitals move to higher energy, indicated by black arrows. Thus, the $p_z$ electrons in BiSe become more unstable than those without strain, which enhances the charge transfer of Bi  $p_z$ electrons to neighboring layers in both upper and lower interfaces.

\section{\label{sec:conclusion} CONCLUSION}

In this work, we systematically study a series of Bi$_2$Se$_3$/BiSe/TMDC heterostructures with small lattice mismatch using first-principles calculations.  With detailed analysis of the band alignment and charge transfer, we show that the rock-salt BiSe buffer layer can donate electrons to both Bi$_2$Se$_3$ and TMDC layers and the magnitude depends on the work function (or electron affinity) difference. We show that TMDCs with a large work function is more favorable to stabilize the BiSe buffer layer and cause a larger amount of charge transfer on the BiSe/TMDC interface, but with a much weaker impact on the charge transfer at the Bi$_2$Se$_3$/BiSe interface. 

The resulting dipole at the Bi$_2$Se$_3$/BiSe interface creates an out-of-plane electric field that breaks the inversion symmetry in the Bi$_2$Se$_3$ layer and introduces an energy offset to the top and bottom Dirac surface states. As a result, new topological properties emerge in few-QL Bi$_2$Se$_3$, including giant Rahsba band splitting in the quantum well states in 2QL-Bi$_2$Se$_3$/BiSe/NbSe$_2$ and the re-appearance of the Dirac point in 3QL-Bi$_2$Se$_3$/BiSe/NbSe$_2$. These results are in excellent agreement with the experimental band structure in the literature from ARPES measurements. In addition, we find that tensile strain can significantly enhance the charge transfer on both interfaces from the BiSe $p_z$ anti-bonding orbitals, which provides another mechanism to control the topological properties of Bi$_2$Se$_3$/BiSe/TMDC heterostructures. 

The emergence of Dirac surface states in quasi-2D materials may open a new avenue for new quantum material platforms, which require the interplay of topological states with strong proximity effects of superconductivity or charge density wave phase.

\begin{acknowledgments}
We thank Dr. Cui-Zu Chang for providing the ARPES data used in this study. This research used resources from the Center for Functional Nanomaterials (CFN), which is a U.S. Department of Energy Office of Science User Facility, at Brookhaven National Laboratory under Contract No. DE-SC0012704. This research also used resources of the National Synchrotron Light Source II, a US Department of energy (DOE) Office of Science User Facility operated for the DOE Office of Science by Brookhaven National Laboratory under Contract number DE-SC0012704.
\end{acknowledgments}




\bibliography{BiSe}

\end{document}